# Dynamics of electrochemical flows III: Closure models

Chengjun Xu[1*], Chin-Tsau Hsu[1,2]

[1]Graduate School at Shenzhen, Tsinghua University, Shenzhen City, Guangdong Province, 518055, China. E-mail address: vivaxuchengjun@163.com

[2]Department of Mechanical Engineering, Hong Kong University of Science and Technology, Kowloon, Hong Kong. E-mail address: mecthsu@ust.hk.

**Abstract:** The electrolyte (comprising of solute ions and solvents) flow-through the porous media is frequently encountered in nature or in many engineering applications, such as the electrochemical systems, manufacturing of composites, oil production, geothermal engineering, nuclear thermal disposal, soil pollution. Our previous work derived the interfacial interaction terms between the solid and the fluid, which can be used to investigate the details of transports of mass, heat, electric flied, potential, or momentum in the process of the electrochemical flows-through porous electrode. In this work, we establish the closure models for these interfacial interaction terms to close the governing equations from mathematical algebra. The interfacial interaction terms regard to the electric field, potential and electric force are firstly revealed. Our new theory provides a new approach to describe the electrochemical flows-through porous media.

## 1. Introduction

The fluid (liquid or gas) flow-through the porous media is frequently encountered in nature or in many engineering applications, such as the electrochemical systems, manufacturing of composites, oil production, geothermal engineering, nuclear thermal disposal, soil pollution, to name a few. Electrochemical systems, for example, electrochemical reactors, batteries, supercapacitors, fuel cells, etc, are widely used in modern society. The energy crisis put the electrochemical conversion and storage systems into the center of today's scientific research around the world. The electrochemical systems generally involve a common and fundamental process, in which the electrolyte (or gas for gas electrode) fluid flows through the porous electrode. The motion of the electrolyte in the electrochemical systems leads to the flows of solvent and ions, known as electrochemical flows. The porous electrode provides a very lager surface area, on which the electrochemical reactions and electrochemical double layer occur, to generate a large current. The improvement and optimum of the electrochemical systems rely on the mathematical modeling or general understanding of the mechanism occurring in such systems [1-6].

In our previous works, we established the general theory for the electrochemical flows-through porous media from the basic conservation laws of charge, mass, momentum, energy and mass concentration. We use static method and set up two representative elementary volumes (REVs). One is the macroscopic REV of the mixture of the porous media and the electrolyte, while the other is the microscopic REV in the electrolyte fluid [7,8]. The macroscopic governing equations are derived by performing the intrinsic volume average on the microscopic equations. After doing that, the terms of dispersion, tortuosity, and solid-fluid interfacial transfer of mass, heat, electric field, and electric force are emerged in the macroscopic equations as the unknown terms. The unknown terms are more than the number of equations. Therefore, it is impossible to solve the equations without the closure modeling. The closure modeling is to construct closure relations based on physics of flows, heat, and mass transfer. The closure modeling has to be got based on the known variables. The closure relations are the links between the unknowns and the known variables.

The closure modeling is based on the dispersed spherical particle assumption. It examines the behaviors of microscopic deviations, for example, $\mathbf{u}'$, $p'$, $T'$ and $C_i'$. The closure modeling has to identify non-dimensional parameters that govern the behaviors of the microscopic transports so as to get the correlation between microscopic deviations and macroscopic quantities. In addition, the closure modeling constructs closure relations applicable in different non-dimensional parameter regimes and composite closure relations valid for all regimes of the parameters.

In this work, we establish the closure models for the dynamics of electrochemical flows-through porous media. Few of the unknown terms in porous media, for example dispersion terms of momentum, heat and mass, interfacial transfer terms of heat and mass, and tortuosity of heat, were well established by previous works [9-13]. However, the unknown terms, with regard to the electric field, potential and electric force, are firstly revealed in this paper. We derive the closure relations for unknown terms with regard to the electric field, potential and electric force from mathematical algebra. The macroscopic equations with the closure models are governing equations of the process of electrochemical flows-through porous media. It is suitable for any kind of the categories of the porous media.

## 2 Closure relations of momentum

The velocity dispersion is

$$\mathbf{u}' = c_1'\bar{\mathbf{u}} + d_p \boldsymbol{c}_1' \cdot \bar{\mathbf{S}} \qquad (1)$$

where $c_1$' and $\boldsymbol{c}_2$' are closure coefficients. Therefore the momentum dispersion is

$$-\overline{\mathbf{u}'\mathbf{u}'} = -c\bar{\mathbf{u}}\bar{\mathbf{u}} + \chi\bar{\mathbf{S}} \qquad (2)$$

χ is the dispersion viscosity and *c* is the dispersion coefficient.

$$\chi = c_1\bar{l}|\bar{\mathbf{u}}| + c_2\bar{l}^2|\bar{\mathbf{S}}| \qquad (3)$$

where $\bar{l}$ is the mixing length of momentum dispersion. According to the VAN-Driest mixing length model, the mixing length is

$$\bar{l} = c_3 d_p \left[1 - \exp\left(-\frac{A^+\bar{x}_3}{d_p}\right)\right] \qquad (4)$$

where *c*, $c_1$, $c_2$ and $c_3$ are closure coefficients and $A^+$ is normally equal to 26. $c_3$ can be taken as 1 by absorbing into $c_1$ and $c_2$.

The interfacial force for integral electrolyte in composite expression is

$$\bar{\mathbf{b}} = -\frac{\phi_1\mu c_S}{d_p^2}\bar{\mathbf{u}} - \frac{\phi_1 c_B}{d_p^{\frac{3}{2}}}\sqrt{\rho_m\mu|\bar{\mathbf{u}}|}\bar{\mathbf{u}} - \frac{\rho_m\phi_1 c_I}{d_p}|\bar{\mathbf{u}}|\bar{\mathbf{u}}$$

$$-\frac{\phi_1 c_G}{d_p}\sqrt{\frac{\rho_m\mu}{|\bar{\nabla}\times(\phi\bar{\mathbf{u}})|}}\bar{\mathbf{u}}\times[\bar{\nabla}\times(\phi\bar{\mathbf{u}})] - \rho_m\phi_1 c_L\bar{\mathbf{u}}\times[\bar{\nabla}\times(\phi\bar{\mathbf{u}})]$$

$$-\sqrt{\frac{\rho_m\mu}{\pi}}\frac{\phi_1 c_M}{d_p}\int_{-\infty}^{t}\frac{\partial(\phi\bar{\mathbf{u}})}{\partial\tau}\frac{d\tau}{\sqrt{t-\tau}} - \rho_m\phi_1 c_V\frac{\bar{D}(\phi\bar{\mathbf{u}})}{\bar{D}t} \qquad (5)$$

where $c_S$ is the drag coefficient due to the Stoke's law, $c_B$ and $c_I$ are viscous and inviscid drag coefficient due to advection, $c_G$ and $c_L$ are viscous and inviscid lift coefficient due to advection, and $c_M$ and $c_V$ are Basset transient memory effect and virtual mass transient inertia drag coefficients. τ is the Basset transient memory effect or virtual mass transient time [8,10].
$\bar{\mathbf{b}}$ depends on the local Reynolds number ($\mathrm{Re_p}$).

$$\mathrm{Re_p} = \frac{|\bar{\mathbf{u}}| d_p}{\nu} \qquad (6)$$

where $\nu$ is the kinetic viscosity.

The selection of terms in interfacial force is based on the given circumstance. For the steady flow, the first three terms are enough to deal with general questions. With a low velocity it only consider the first term, while for the high velocity the second and third term will take over. While if the flow is transient flow, for example discharging a battery with pulse, the Basset transient memory effect and virtual force have to be considered.

3 Closure relations of heat transfer

The closure model for the heat conduction was well established by C.T. Hsu. However, in his model there is no heat generated at the interface between the fluid and the solid. In the electrochemical systems there generally has the electrochemical reaction to generate or absorb the heat at the interface. Therefore, the heat flux from the solid to the fluid will be influenced by the heat generated at the interface. If there is no heat generated at the interface, the closure model of C.T. Hsu can be directly used.

Let’s assume that the porous material consists of randomly packed solid particles surrounded by fluids. For simplicity, we take the uniform diameter of $d_p$ to characterize microscopic scale of the media. We also assume that the particle size of $d_p$ is much larger than the typical size of molecules such that the fluid and the solid microscopically are regarded as continuum. Hence the microscopic transient heat conduction equations for the fluid and solid are given by

$$\rho_m C_p \frac{\partial T}{\partial t} = -\nabla \cdot (k \nabla T) \qquad (7)$$

$$\rho_{m1} C_{p1} \frac{\partial T_1}{\partial t} = -\nabla \cdot (k_1 T_1) \qquad (8)$$

where the subscripts 1 refers to the solid. Not that we can easily find a system that is no heat convection. The proper boundary conditions on the fluid-solid interface of $A_{12}$ are

$$T = T_1 \qquad on\ A_{12} \qquad (9)$$

and

$$\boldsymbol{n} \cdot k\nabla T = \boldsymbol{n} \cdot k\nabla T_1 + \boldsymbol{n} \cdot \mathrm{H}_E \qquad on\ A_{12} \qquad (10)$$

where $\boldsymbol{n}$ is the unit vector out normal from fluid to solid and $\mathrm{H}_E$ is the rate of heat generated at the interface.

It is impractical to solve above four equations with details, especially when the number of solid particles are large. Alternatively, we are more interested in the global characteristics of heat conduction in porous media. To this end, we consider the macroscopic equations for heat conduction

$$\rho_m C_p \frac{\partial(\phi \bar{T})}{\partial t} = -\nabla \cdot (k(\phi \bar{T})) + k\bar{\nabla} \cdot \Lambda_{12} + \mathbf{q}_{12} \qquad (11)$$

$$\rho_{m1} C_{p1} \frac{\partial[(1-\phi)\bar{T}_1]}{\partial t} = -\nabla \cdot [k_1((1-\phi)\bar{T}_1)] - k_1\bar{\nabla} \cdot \Lambda_{12} - \mathbf{q}_{12} \qquad (12)$$

To close the equations 11 and 12, we need to constitute equations, which relate the integral terms to the macroscopically phase-averaged temperatures, $\bar{T}$ and $\bar{T}_1$. To this end, we first decompose the temperature into

$$T = \bar{T} + T' \qquad (13)$$

$$T_1 = \bar{T}_1 + T_1' \qquad (14)$$

where $T_1'$ and $T'$ represent the microscopic temperature variations from the phase-averaged values. As the macroscopic REV length scale of $l$ is much larger than the particle size of $d_p$, the time scale of macroscopic conduction is $l^2/\alpha$, which is much larger than the time scale of microscopic conduction of $d_p{}^2/\alpha$. Therefore, with respect to macroscopic process of long time scale, the local microscopic heat conduction process is quasi-steady. The microscopic equations become by invoking the equations 13 and 14.

$$\rho_m C_p [\frac{\partial \bar{T}}{\partial t} + \frac{\partial T'}{\partial t}] = -\nabla \cdot (k\nabla \bar{T}) - \nabla \cdot (k\nabla T') \qquad (15)$$

$$\rho_{m1}C_{p1}[\frac{\partial \bar{T}_1}{\partial t}+\frac{\partial T_1'}{\partial t}] = -\nabla\cdot(k_1\nabla\bar{T}_1) - \nabla\cdot(k_1\nabla T_1') \qquad (16)$$

Under the quasi-steady assumption, the dispersion terms will vanish and we get

$$\nabla\cdot(k\nabla T') = 0 \qquad (17)$$

$$\nabla\cdot(k_1\nabla T_1') = 0 \qquad (18)$$

The interfacial boundary conditions now become

$$T' = T_1' + (\bar{T}_1 - \bar{T}) \quad on\ A_{12} \qquad (19)$$

$$\boldsymbol{n}\cdot T' = \boldsymbol{n}\cdot\sigma_h\nabla T_1' + \boldsymbol{n}\cdot[\sigma_h\nabla\bar{T}_1 - \nabla\bar{T}] + \boldsymbol{n}\cdot\frac{\mathrm{H}_E}{k} \qquad on\ A_{12} \qquad (20)$$

where $\sigma_h$ is the thermal conductivity ratio between solid and fluid.

$$\sigma_h = \frac{k_1}{k} \qquad (21)$$

The general solutions of $T_1'$ and $T'$ will take the form of

$$T' = f_0'(\bar{T}_1 - \bar{T}) + \mathbf{f}_1'\cdot[\nabla\bar{T} - \sigma_h\nabla\bar{T}_1] + \mathbf{f}_2'\cdot\frac{\mathrm{H}_E}{k} \qquad (22)$$

and

$$T_1' = g_0'(\bar{T}_1 - \bar{T}) + \mathbf{g}_1'\cdot[\nabla\bar{T} - \sigma_h\nabla\bar{T}_1] + \mathbf{g}_2'\cdot\frac{\mathrm{H}_E}{k} \qquad (23)$$

According to the model of Hsu and Cheng, the thermal dispersion is

$$-\phi\overline{\mathbf{u}'T'} = \mathbf{A}_\mathrm{D}\cdot\bar{\nabla}\bar{T} \qquad (24)$$

$\mathbf{A}_\mathrm{D}$ is the thermal dispersion diffusivity tensor. For the homogeneous media, we get

$$\mathbf{A}_{\mathrm{D}} = \begin{bmatrix} \alpha_1 & 0 & 0 \\ 0 & \alpha_2 & 0 \\ 0 & 0 & \alpha_3 \end{bmatrix} \qquad (25)$$

where $\alpha_1$, $\alpha_2$ and $\alpha_3$ are the thermal dispersion diffusivities in the longitudinal, transverse and lateral directions, respectively [11].

The thermal dispersion diffusivity depends on local Peclet number ($\mathrm{Pe_p}$)

$$\mathrm{Pe_P} = \frac{|\bar{\mathbf{u}}| d_p}{\alpha} \qquad (26)$$

When the Peclet number is high the thermal dispersion diffusivity is linearly proportional to Peclet number, while when the Peclet number is low the thermal dispersion diffusivity is linearly proportional to the square of Peclet number. The composite expression for the dispersion thermal diffusivities is

$$\alpha_i = (1-\phi)\alpha \frac{a_i \mathrm{Pe_P^2}}{b_i + \mathrm{Pe_P}} \quad (i = 1, 2 \; and \; 3) \quad (27)$$

where $a_i$ and $b_i$ are coefficients.

The thermal tortuosity is

$$\Lambda_{12} = \boldsymbol{G}_0(\bar{T}_1 - \bar{T}) + \boldsymbol{G_1} \cdot (\bar{\nabla}\bar{T} - \sigma_h \bar{\nabla}\bar{T}_1) + \boldsymbol{G_2} \cdot \bar{\mathrm{H}}_E \qquad (28)$$

where

$$\boldsymbol{G}_0 = \frac{1}{V} \int_{A_{12}} f_0' \, d\boldsymbol{s} \qquad (29a)$$

$$\boldsymbol{G}_1 = \frac{1}{V} \int_{A_{12}} \mathbf{f}_1' \cdot d\boldsymbol{s} \qquad (29b)$$

$$\boldsymbol{G}_2 = \frac{1}{V} \int_{A_{12}} \frac{\mathbf{f}_2'}{k} \cdot d\boldsymbol{s} \qquad (29c)$$

$$\mathbf{q}_{12} = \boldsymbol{M}_0(\bar{T}_1 - \bar{T}) + \boldsymbol{M}_\mathbf{1} \cdot (\bar{\nabla}\bar{T} - \sigma_h \bar{\nabla}\bar{T}_1) + \boldsymbol{M}_2 \cdot \bar{\mathrm{H}}_E \qquad (30)$$

where

$$\boldsymbol{M}_0 = \frac{1}{V}\int_{A_{12}} k\nabla f_0' \, d\boldsymbol{s} \qquad (31a)$$

$$\boldsymbol{M}_1 = \frac{1}{V}\int_{A_{12}} k\nabla \mathbf{f}_1' \cdot d\boldsymbol{s} \qquad (31b)$$

$$\boldsymbol{M}_2 = \frac{1}{V}\int_{A_{12}} \nabla \mathbf{f}_2' \cdot d\boldsymbol{s} \qquad (31c)$$

For randomly distributed particles, the volumetric average tensor $\boldsymbol{G}_1$ has to be axially symmetric to exhibit an isotropic properties, i.e., $\boldsymbol{G}_1 = G\boldsymbol{I}$, where $\boldsymbol{I}$ is the unit tensor matrix and $G$ is the scalar quantity. According to Newton's Law, the local interfacial heat transfer is proportional to the total fluid-solid interfacial area $A_{12}$. Therefore, we have

$$\boldsymbol{M}_0 = h_{12} a_{12} \qquad (32a)$$

$$\boldsymbol{M}_2 = M_{12} a_{12} \qquad (32b)$$

$a_{12}$ is the interfacial area density.

$$a_{12} = \frac{A_{12}}{V} \qquad (33)$$

$h_{12}$ and $M_{12}$ are the interfacial heat transfer coefficients.

From the physical point of view, a nonzero value in $\boldsymbol{G}_0$ and $\boldsymbol{M}_1$ will lead to a convective behavior associated with the macroscopic temperature gradient, where vectors $\boldsymbol{G}_0$ and $\boldsymbol{M}_1$ resemble the velocities of convection. Hsu and Quintard had argued that this convection behavior does not have a physical ground since all these transfer process are essentially from conduction. They also demonstrated mathematically that $\boldsymbol{G}_0 = \boldsymbol{M}_1 = 0$. In addition, in the thermal tortuosity term, the nonzero value of $\boldsymbol{G}_2$ will lead to the convection

Therefore, thermal tortuosity is

$$\Lambda_{12} = \boldsymbol{G}(\overline{\nabla}\bar{T} - \sigma_h \overline{\nabla}\bar{T}_1) \qquad (34)$$

The interfacial heat transfer is

$$\mathbf{q}_{12} = h_{12}a_{12}(\bar{T}_1 - \bar{T}) + M_{12}a_{12} \cdot \overline{\mathrm{H}}_E \qquad (35)$$

where $\overline{\mathrm{H}}_E$ is the intrinsic heat generated at the interface.

$$\overline{\mathrm{H}}_E = \omega\bar{\mathbf{J}}_E \qquad (36)$$

At the thermal equilibrium condition, we get

$$\bar{T}_1 = \bar{T} \qquad (37)$$

Equations 11 and 12, after invoking equation 37, are added together to get the heat conduction equation for the solid-fluid mixture.

$$(\rho_m C_p)_m \frac{\partial \bar{T}}{\partial t} = -\overline{\nabla} \cdot (k_{st}\overline{\nabla}\bar{T}) \qquad (38)$$

where

$$(\rho_m C_p)_m = \phi\rho_m C_p + (1-\phi)\rho_{m1}C_{p1} \qquad (39a)$$

$$k_{st} = \phi k + (1-\phi)k_1 + k(1-\sigma_h)^2\boldsymbol{G} \qquad (39b)$$

Therefore, $\boldsymbol{G}$ is the thermal tortuosity parameter.

$$\boldsymbol{G} = \frac{\frac{k_{st}}{k} - \phi - \sigma_h(1-\phi)}{(1-\sigma_h)^2} \qquad (40)$$

$k_{st}$ is the effective stagnant thermal conductivity under the condition of local thermal equilibrium. Steady state effective thermal conductivity of the media is also considered as a composite.

The interfacial heat transfer coefficient is

$$h_{12} = h_{12}^{*}(1 + c_4 Pr \mathrm{Re_P}) \qquad (41)$$

When the local Reynold number ($\mathrm{Re_P}$) is low, the $h^{*}_{12}$ is the stagnant interfacial heat transfer coefficient. $c_4$ is the coefficient. $Pr$ is the Prandtl number [12].

4 Closure relations of mass transfer

The concentration conservation is similar to the energy conservation. The dispersion of the mass concentration, mass tortuosity and interfacial mass transfer as well as dispersion and tortuosity of electric force cloud be derived from the mass conservation equation similar to the closure model of heat transfer.

Let's assume that the porous material consists of randomly packed solid particles surrounded by fluids. We also assume that the particle size of $d_p$ is much larger than the typical size of molecules such that fluid and solid microscopically are regarded as continuum. Hence the microscopic transient mass diffusion and conductance equations in the macroscopic REV for the fluid and solid are given by

$$\frac{\partial C_i}{\partial t} = \nabla \cdot (D_i \nabla C_i) + \nabla \cdot (\gamma C_i \nabla \mathbf{\Phi}_2) \qquad (42)$$

$$\frac{\partial C_{1i}}{\partial t} = \nabla \cdot (D_{1i} \nabla C_{1i}) + \nabla \cdot (\gamma_1 C_{1i} \nabla \mathbf{\Phi}_1) \qquad (43)$$

where $C_i$ is the concentration of ion species $i$, $D_i$ is the diffusion diffusivity coefficient, the subscripts 1 refers to the solid phase. $\gamma$ and $\gamma_1$ are the coefficient at the thermal equilibrium.

$$\gamma = \frac{z_i F}{RT} D_i \quad (44)$$

$$\gamma_1 = \frac{z_i F}{RT} D_{1i} \quad (45)$$

Not that we can easily find a system that is no convection of mass transfer. The proper boundary conditions on the fluid-solid interface of $A_{12}$ are

$$C_{1i} = C_i \qquad on\ A_{12} \qquad (46a)$$

$$\boldsymbol{n} \cdot D_{1i} \nabla C_{1i} = \boldsymbol{n} \cdot D_i \nabla C_i \qquad on\ A_{12} \qquad (46b)$$

and

$$\boldsymbol{\Phi}_1 = \boldsymbol{\Phi} = \boldsymbol{\Phi}_2 + \eta + \boldsymbol{\Phi}_d \qquad on\ A_{12} \qquad (46c)$$

$$\boldsymbol{n} \cdot \gamma_1 C_{1i} \nabla \boldsymbol{\Phi}_1 = \boldsymbol{n} \cdot \gamma C_i \nabla \boldsymbol{\Phi} \qquad on\ A_{12} \qquad (46d)$$

where $\boldsymbol{n}$ is the unit vector out normal from fluid to solid, $\eta$ is the overpotential and $\boldsymbol{\Phi}_d$ is the potential of the electrochemical double layer.
We first decompose the concentration and potential into

$$C_i = \bar{C}_i + C_i' \qquad (47a)$$

$$C_{1i} = \bar{C}_{1i} + C_{1i}' \qquad (47b)$$

$$\boldsymbol{\Phi}_1 = \bar{\boldsymbol{\Phi}}_1 + \boldsymbol{\Phi}_1' \qquad (47c)$$

$$\boldsymbol{\Phi} = \bar{\boldsymbol{\Phi}} + \boldsymbol{\Phi}' = \bar{\boldsymbol{\Phi}}_2 + \bar{\eta} + \bar{\boldsymbol{\Phi}}_{\boldsymbol{d}} + \boldsymbol{\Phi}_2' \qquad (47d)$$

where $C_{1i}'$ and $C_i'$ represent the microscopic concentration variations in solid and fluid from the phase-averaged values, respectively. As the macroscopic REV length scale of $l$ is much larger than the particle size of $d_p$, the time scale of macroscopic conduction is $l^2/D_i$, which is much larger than the time scale of microscopic conduction of $d_p{}^2/D_i$. Therefore, with respect to macroscopic process of long time scale, the local microscopic diffusion and conduction process is quasi-steady. The microscopic equations become by invoking the equation 47.

$$\frac{\partial \bar{C}_i}{\partial t}+\frac{\partial C_i'}{\partial t}= \nabla\cdot(D_i\nabla\bar{C}_i)+\nabla\cdot(\gamma\bar{C}_i\nabla\overline{\mathbf{\Phi}})+\nabla\cdot(D_i\nabla C_i')+\nabla\cdot(\gamma C_i'\nabla\mathbf{\Phi}') \qquad (48)$$

$$\frac{\partial \bar{C}_{1i}}{\partial t}+\frac{\partial C_{1i}'}{\partial t}= \nabla\cdot(D_{1i}\nabla\bar{C}_{1i})+\nabla\cdot(\gamma_1\bar{C}_{1i}\nabla\overline{\mathbf{\Phi}}_1)+\nabla\cdot(D_{1i}\nabla C_{1i}')+\nabla\cdot(\gamma_1 C_{1i}'\nabla\mathbf{\Phi}_2') \quad (49)$$

Under the quasi-steady assumption, the dispersion term will vanish. Considering the mass diffusion we get

$$\nabla\cdot(D_i\nabla C_i')=0 \qquad (50a)$$

$$\nabla\cdot(D_{1i}\nabla C_{1i}')=0 \qquad (50b)$$

The interfacial boundary conditions now are

$$C_i' = C_{1i}'+(\bar{C}_{1i}-\bar{C}_i) \quad on\ A_{12} \qquad (51a)$$

$$\boldsymbol{n}\cdot C_i' = \boldsymbol{n}\cdot\sigma_{mi}\nabla C_{1i}'+\boldsymbol{n}\cdot[\sigma_{mi}\nabla\bar{C}_{1i}-\nabla\bar{C}_i] \qquad on\ A_{12} \qquad (51b)$$

where $\sigma_{mi}$ is the mass diffusion coefficient ratio between solid and fluid for ion species *i*.

$$\sigma_{mi}=\frac{D_{1i}}{D_i} \qquad (52)$$

The general solutions of $C_{1i}'$ and $C_i'$ will take the form of

$$C_i' = f_{m0}'(\bar{C}_{1i}-\bar{C}_i)+\mathbf{f}_{m1}'\cdot[\nabla\bar{C}_i-\sigma_{mi}\nabla\bar{C}_{1i}] \qquad (53a)$$

and

$$C_{1i}' = g_{m0}'(\bar{C}_{1i}-\bar{C}_i)+\mathbf{g}_{m1}'\cdot[\nabla\bar{C}_i-\sigma_{mi}\nabla\bar{C}_{1i}] \qquad (53b)$$

The mass dispersion is

$$-\phi\overline{C_i'\mathbf{u}_i'} = \mathbf{M}_{\mathrm{D}i} \cdot \overline{\nabla}\bar{C}_i \qquad (54)$$

where $\mathbf{M}_{\mathrm{D}i}$ is the mass dispersion diffusion coefficient tensor of ion species $i$. For homogeneous media, we get

$$\mathbf{M}_{\mathrm{D}i} = \begin{bmatrix} D_{i1} & 0 & 0 \\ 0 & D_{i2} & 0 \\ 0 & 0 & D_{i3} \end{bmatrix} \qquad (55)$$

where $D_{i1}$, $D_{i2}$ and $D_{i3}$ are the mass dispersion diffusivity coefficients in the longitudinal, transverse and lateral directions, respectively.

The mass diffusion coefficient is

$$D_i = \frac{\lambda_i}{\rho_m} \qquad (56)$$

The composite expression for the mass dispersion diffusion coefficients is

$$D_{ij} = (1-\phi)\,\frac{a_{mi}(Pe_{mp})^2}{b_{mi} + Pe_{mp}} \quad (j = 1, 2 \text{ and } 3) \qquad (57)$$

where $a_{mi}$ and $b_{mi}$ are the coefficients and $Pe_{mp}$ is the local mass transfer Peclet number.

$$Pe_{mp} = \frac{|\bar{\mathbf{u}}_i| d_p}{D_i} \qquad (58)$$

At low mass transfer Peclet number, the mass dispersion coefficient is linearly proportional to mass transfer Peclet number, while the mass dispersion diffusivity is proportional to square of mass transfer Peclet number when the mass Peclet number is high [13].

Mass diffusion tortuosity parameter is very similar to thermal tortuosity parameter.

$$\mathbf{G}_{mi} = \frac{\frac{D_{sti}}{D_i} - \phi - \sigma_{mi}(1-\phi)}{(1-\sigma_{mi})^2} \qquad (59)$$

where $D_{1i}$ are $D_i$ the diffusion coefficients of species $i$ in solid and fluid, respectively.

The mass transfer tortuosity is

$$\mathbf{\Omega}_i = \mathbf{G}_{mi}\overline{\nabla}\bar{C}_i \qquad (60)$$

Under the condition of $\sigma_{mi} << 1$, $\mathbf{G}_{mi}$ reduces to

$$\mathbf{G}_{mi} = \frac{D_{sti}}{D_i} - \phi \qquad (61)$$

$D_{sti}$ is effective stagnant mass diffusion coefficient in fluid for species $i$ under the condition of local thermal equilibrium, also steady state effective mass coefficient.

The interfacial mass transfer is defined as

$$m_{12} = g_{12}a_{12}(C_i^0 - \bar{C}_i) \qquad (62)$$

where $g_{12}$ is interfacial mass transfer coefficient, $a_{12}$ is the interfacial area density, and $C_i^0$ is the stagnant (bulk) concentration of species $i$ in the fluid.

When the Reynolds number is low, the interfacial transfer coefficient is

$$g_{12} = g_{12}^{*}(1 + C_5 Sc\mathrm{Re_p}) \qquad (63)$$

where $g^{*}_{12}$ is the stagnant interfacial mass transfer coefficient, $c_5$ is the coefficient, and $Sc$ is Schmidt number.

$$Sc = \frac{\mu}{\lambda_i} \qquad (64)$$

where $\mu$ is the dynamic viscosity of the fluid and $\lambda_i$ is the diffusion conductivity of species $i$.

The composite expression for interfacial mass transfer can be expressed in term of Sherwood number ($Sh$).

$$Sh = \frac{g_{12}d_p}{\lambda_i} = Sh^{*}(1 + \frac{a_{mi}Sc\mathrm{Re_p}}{b_{mi} + Sc^{2/3}\mathrm{Re_p^{1-n}}}) \qquad (65)$$

where $Sh^*$ is stagnant interfacial Sherwood number and $n$ depends on the ranges of $Re_p$.

$$Sh^* = \frac{g_{12}^* d_p}{\lambda_i} \quad (66)$$

$a_{mi}$ and $b_{mi}$ are coefficients.

$$c_5 = \frac{a_{mi}}{b_{mi}} \quad (67)$$

5 Closure relations of electric force

Because $C'_{1i}$ and $C'_i$ is known to be expressed as the function of $\bar{C}_{1i}$ and $\bar{C}_i$, under the quasi-steady assumption, by only considering the mass conduction by electric force we get

$$\nabla \cdot (\gamma' \nabla \boldsymbol{\Phi}') = 0 \quad (68a)$$

$$\nabla \cdot (\gamma'_1 \nabla \boldsymbol{\Phi}'_1) = 0 \quad (68\mathrm{b})$$

where

$$\gamma' = \gamma C'_i = \frac{z_i F}{RT} D_i C'_i \quad (69a)$$

$$\gamma'_1 = \gamma_1 C'_{1i} = \frac{z_i F}{RT} D_{1i} C'_{1i} \quad (69b)$$

The interfacial boundary conditions now become

$$\boldsymbol{\Phi}'_2 = \boldsymbol{\Phi}'_1 + (\overline{\boldsymbol{\Phi}}_1 - \overline{\boldsymbol{\Phi}}) \quad on\ A_{12} \qquad (70a)$$

$$\boldsymbol{n} \cdot \boldsymbol{\Phi}' = \boldsymbol{n} \cdot \sigma_{ei} \nabla \boldsymbol{\Phi}'_1 + \boldsymbol{n} \cdot (\sigma_{ei} \nabla \overline{\boldsymbol{\Phi}}_1 - \nabla \overline{\boldsymbol{\Phi}}) \quad on\ A_{12} \qquad (70b)$$

where $\sigma_{ei}$ is the electric force coefficient ratio between solid and fluid for ion species $i$.

$$\sigma_{ei} = \frac{\gamma'_1}{\gamma'} = \frac{D_{1i} C'_{1i}}{D_i C'_i} = \sigma_{mi} \qquad (71)$$

The general solutions of $\mathbf{\Phi}_1'$ and $\mathbf{\Phi}'$ will take the form of

$$\mathbf{\Phi}' = f_{e0}'(\bar{\mathbf{\Phi}}_1 - \bar{\mathbf{\Phi}}) + \mathbf{f}_{e1}' \cdot (\nabla\bar{\mathbf{\Phi}} - \sigma_{ei}\nabla\bar{\mathbf{\Phi}}_1) \qquad (72)$$

and

$$\mathbf{\Phi}_1' = g_{e0}'(\bar{\mathbf{\Phi}}_1 - \bar{\mathbf{\Phi}}) + \mathbf{g}_{e1}' \cdot (\nabla\bar{\mathbf{\Phi}} - \sigma_{ei}\nabla\bar{\mathbf{\Phi}}_1) \qquad (73)$$

Therefore, we get the gradient of potential dispersion

$$\nabla\mathbf{\Phi}' = \nabla f_{e0}'(\bar{\mathbf{\Phi}}_1 - \bar{\mathbf{\Phi}}) + \nabla\mathbf{f}_{e1}' \cdot (\nabla\bar{\mathbf{\Phi}} - \sigma_{ei}\nabla\bar{\mathbf{\Phi}}_1) \qquad (74)$$

$$\nabla\mathbf{\Phi}_1' = \nabla g_{e0}'(\bar{\mathbf{\Phi}}_1 - \bar{\mathbf{\Phi}}) + \nabla\mathbf{g}_{e1}' \cdot (\nabla\bar{\mathbf{\Phi}} - \sigma_{ei}\nabla\bar{\mathbf{\Phi}}_1) \qquad (75)$$

The dispersion of the force normally is equal to a dispersion coefficient plus the force. For example, the dispersion of electric force in the electrolyte is

$$-\phi\gamma\overline{C_i'\nabla\mathbf{\Phi}_2'} = \gamma_d\bar{C}_i\bar{\nabla}\bar{\mathbf{\Phi}}_2 \qquad (76)$$

where $\gamma_d$ is the dispersion coefficient. Looking at the coefficient $\gamma$, at the thermal equilibrium its dispersion can only come from the dispersion of diffusion coefficient. Therefore, the dispersion of electric force is

$$-\phi\overline{C_i'\nabla\mathbf{\Phi}_2'} = \mathrm{F}_{Di} \cdot \bar{C}_i\bar{\nabla}\bar{\mathbf{\Phi}}_2 \qquad (77a)$$

$$-\phi\gamma\overline{C_i'\nabla\mathbf{\Phi}_2'} = \mathbf{F}_{\mathrm{D}i} \cdot \bar{C}_i\bar{\nabla}\bar{\mathbf{\Phi}}_2 \qquad (77b)$$

$$-\phi_1\gamma_1\overline{C_{1i}'\nabla\mathbf{\Phi}_1'} = \mathbf{F}_{1\mathrm{D}i} \cdot \bar{C}_{1i}\bar{\nabla}\bar{\mathbf{\Phi}}_1 \qquad (77c)$$

where $\mathrm{F}_{\mathrm{D}i}$, $\mathbf{F}_{\mathrm{D}i}$, and $\mathbf{F}_{\mathrm{D}1i}$ are the electric force dispersion coefficients of species $i$. For homogeneous media, we get

$$\mathrm{F}_{\mathrm{D}i} = \begin{bmatrix} \frac{D_{i1}}{D_i} & 0 & 0 \\ 0 & \frac{D_{i2}}{D_i} & 0 \\ 0 & 0 & \frac{D_{i3}}{D_i} \end{bmatrix} \qquad (78a)$$

$$\mathbf{F}_{\mathrm{D}i} = \begin{bmatrix} \frac{z_i F D_{i1}}{RT} & 0 & 0 \\ 0 & \frac{z_i F D_{i2}}{RT} & 0 \\ 0 & 0 & \frac{z_i F D_{i3}}{RT} \end{bmatrix} \qquad (78b)$$

$$\mathbf{F}_{1\mathrm{D}i} = \begin{bmatrix} \frac{z_i F D_{1i1}}{RT} & 0 & 0 \\ 0 & \frac{z_i F D_{1i2}}{RT} & 0 \\ 0 & 0 & \frac{z_i F D_{1i3}}{RT} \end{bmatrix} \qquad (78c)$$

where $D_{1i1}$, $D_{1i2}$ and $D_{1i3}$ are the mass dispersion diffusion coefficients of species $i$ in solid in the longitudinal, transverse and lateral directions, respectively.

The tortuosity of the electric force is

$$\mathbf{\Gamma}_i = \frac{1}{V}\int_{A_{12}} C_i \nabla \mathbf{\Phi}_2 \cdot d\boldsymbol{s}$$

$$= \boldsymbol{G}_{e0}(\nabla\bar{C}_i - \sigma_{mi}\nabla\bar{C}_{1i})(\bar{\mathbf{\Phi}}_1 - \bar{\mathbf{\Phi}}) + \boldsymbol{G}_{e1}(\nabla\bar{C}_i - \sigma_{mi}\nabla\bar{C}_{1i}) \cdot (\nabla\bar{\mathbf{\Phi}} - \sigma_{ei}\nabla\bar{\mathbf{\Phi}}_1) \qquad (79)$$

where

$$\boldsymbol{G}_{e0} = \frac{1}{V}\int_{A_{12}} f'_{e0}\,\nabla f'_{e0} \cdot d\boldsymbol{s} = 0 \qquad (80a)$$

and

$$\boldsymbol{G}_{e1} = \frac{1}{V}\int_{A_{12}} f'_{e0}\nabla\cdot\mathbf{f}'_1\, d\boldsymbol{s} \qquad (80b)$$

$\boldsymbol{G}_{e0}$ is equal to zero because the plus of gradient of concentration and potential has no physical meanings. And at the equilibrium, this term will vanish.

Therefore, the electric force tortuosity for species *i* is

$$\boldsymbol{\Gamma}_i = \boldsymbol{G}_{ei}(\overline{\nabla}\bar{C}_i - \sigma_{mi}\overline{\nabla}\bar{C}_{1i})\cdot(\nabla\overline{\boldsymbol{\Phi}} - \sigma_{ei}\nabla\overline{\boldsymbol{\Phi}}_1) \qquad (81)$$

At the equilibrium, the macroscopic equations for the stagnant mass transfer process under conduction and diffusion for solid and fluid are

$$\frac{\partial(\phi\bar{C}_i)}{\partial t} = D_i\overline{\nabla}^2(\phi\bar{C}_i) + D_i\overline{\nabla}\cdot\boldsymbol{\Omega}_i + m_{12} + \gamma\overline{\nabla}\cdot[\bar{C}_i\overline{\nabla}(\phi\overline{\boldsymbol{\Phi}})] + \gamma\overline{\nabla}\cdot\boldsymbol{\Gamma}_i + \gamma\boldsymbol{\Gamma}_i \qquad (82)$$

$$\frac{\partial[(1-\phi)\bar{C}_{1i}]}{\partial t} = D_{1i}\overline{\nabla}^2((1-\phi)\bar{C}_{1i}) + -D_{1i}\overline{\nabla}\cdot\boldsymbol{\Omega}_i - m_{12} + \gamma_1\overline{\nabla}\cdot[\bar{C}_{1i}\overline{\nabla}((1-\phi)\overline{\boldsymbol{\Phi}}_1)] - \gamma_1\overline{\nabla}\cdot\boldsymbol{\Gamma}_i - \gamma_1\boldsymbol{\Gamma}_i \qquad (83)$$

Under the local equilibrium in the macroscopic REV, we get

$$\bar{C}_{1i} = \bar{C}_i \qquad (84)$$

$$\overline{\boldsymbol{\Phi}}_1 = \overline{\boldsymbol{\Phi}} \qquad (85)$$

By summing up equations 82 and 83 and invoking equations 84 and 85, we get

$$\frac{\partial\bar{C}_i}{\partial t} = \overline{\nabla}\cdot[\overline{\nabla}(D_{sti}\bar{C}_i)] + \overline{\nabla}\cdot(\gamma_{st}\bar{C}_i\overline{\nabla}\overline{\boldsymbol{\Phi}}) \qquad (86)$$

where $\gamma_{st}$ is the effective stagnant coefficient.

$$\gamma_{st} = \phi\gamma + (1-\phi)\gamma_1 + \gamma(1-\sigma_{mi})^3\boldsymbol{G}_{ei} \qquad (87)$$

Therefore, the tortuosity parameter ($\boldsymbol{G}_{ei}$) of electric force is

$$\boldsymbol{G}_{ei} = \frac{\frac{\gamma_{st}}{\gamma} - \phi - \sigma_{mi}(1-\phi)}{(1-\sigma_{mi})^3} \quad (88)$$

where $C_{sti}$ is the effective stagnant concentration under the condition of local thermal equilibrium. When $\sigma_{mi}$<< 1, the tortuosity parameter ($\boldsymbol{G}_{ei}$) is

$$\boldsymbol{G}_{ei} = \frac{\gamma_{st}}{\gamma} - \phi \quad (89)$$

After the derivation of tortuosity of heat, mass and electric force from mathematical algebra, it is necessary to discuss the physical meaning of the tortuosity. The expressions of the tortuosity of heat, mass and electric force are

$$k_{st} = \phi k + (1-\phi)k_1 + k(1-\sigma_h)^2\boldsymbol{G} \qquad (90a)$$

$$D_{sti} = \phi D_i + (1-\phi)D_{1i} + D_i(1-\sigma_{mi})^2\mathbf{G}_{mi} \qquad (90b)$$

$$\gamma_{st} = \phi\gamma + (1-\phi)\gamma_1 + \gamma(1-\sigma_{mi})^3\mathbf{G}_{ei} \quad (90c)$$

If there is no tortuosity, the equation becomes

$$k_{st} = \phi k + (1-\phi)k_1 \qquad (91a)$$

$$D_{sti} = \phi D_i + (1-\phi)D_{1i} \qquad (91b)$$

$$\gamma_{st} = \phi\gamma + (1-\phi)\gamma_1 \quad (91c)$$

This means the fluid phase and the solid phase are parallel. In this type of configuration, the transference of the heat, mass and electric force solely occurs in each phase. The appearance of the tortuosity leads to the handicap of the transference of the heat, mass and electric force. For example, the appearance of the tortuosity forces the heat transfer to be sinuous.

In addition the stagnant coefficients are the coefficients when the fluid is not moving. Therefore, there is a relation

$$\alpha = \alpha_{st} + \mathbf{A}_{\mathrm{D}} \qquad (92a)$$

$$D_i = D_{sti} + \mathbf{M}_{\mathrm{D}i} \qquad (92b)$$

$$\gamma = \gamma_{st} + \mathbf{F}_{\mathrm{D}i} \qquad (92c)$$

where $\alpha_{st}$ is stagnant thermal diffusivity.

7. Tortuosity of potential and electric field

The tortuosity for the electric field and potential is

$$\boldsymbol{\Psi}_1 = \frac{1}{V}\int_{\boldsymbol{A}_{12}} \boldsymbol{\Phi}_1 \cdot d\boldsymbol{s} = a_5(\bar{\boldsymbol{\Phi}}_1 - \bar{\boldsymbol{\Phi}}) + \boldsymbol{a}_6 \cdot (\nabla\bar{\boldsymbol{\Phi}} - \sigma_{ei}\nabla\bar{\boldsymbol{\Phi}}_1) \qquad (93a)$$

$$\boldsymbol{\Psi}_2 = \frac{1}{V}\int_{\boldsymbol{A}_{12}} \boldsymbol{\Phi}_2 \cdot d\boldsymbol{s} = a_7(\bar{\boldsymbol{\Phi}}_1 - \bar{\boldsymbol{\Phi}}) + \boldsymbol{a}_8 \cdot (\nabla\bar{\boldsymbol{\Phi}} - \sigma_{ei}\nabla\bar{\boldsymbol{\Phi}}_1) \qquad (93b)$$

$$\boldsymbol{\Xi}_1 = \frac{1}{V}\int_{\boldsymbol{A}_{12}} \nabla\boldsymbol{\Phi}_1 \cdot d\boldsymbol{s} = a_9(\bar{\boldsymbol{\Phi}}_1 - \bar{\boldsymbol{\Phi}}) + \boldsymbol{a}_{10} \cdot (\nabla\bar{\boldsymbol{\Phi}} - \sigma_{ei}\nabla\bar{\boldsymbol{\Phi}}_1) \qquad (93c)$$

$$\boldsymbol{\Xi}_2 = \frac{1}{V}\int_{\boldsymbol{A}_{12}} \nabla\boldsymbol{\Phi}_2 \cdot d\boldsymbol{s} = a_{11}(\bar{\boldsymbol{\Phi}}_1 - \bar{\boldsymbol{\Phi}}) + \boldsymbol{a}_{12} \cdot (\nabla\bar{\boldsymbol{\Phi}} - \sigma_{ei}\nabla\bar{\boldsymbol{\Phi}}_1) \qquad (93d)$$

where $a_5$, $\boldsymbol{a}_6$, $a_7$, $\boldsymbol{a}_8$ ,$a_9$, $\boldsymbol{a}_{10}$, $a_{11}$,and $\boldsymbol{a}_{12}$ and are closure coefficients.

$$a_5 = \frac{1}{V}\int_{\boldsymbol{A}_{12}} g'_{e0} \cdot d\boldsymbol{s} \qquad (94a)$$

$$\boldsymbol{a}_6 = \frac{1}{V}\int_{\boldsymbol{A}_{12}} \mathbf{g}'_{e1} \cdot d\boldsymbol{s} \qquad (94b)$$

$$a_7 = \frac{1}{V}\int_{A_{12}} f'_{e0} \cdot d\mathbf{s} \qquad (94c)$$

$$\boldsymbol{a}_8 = \frac{1}{V}\int_{A_{12}} \mathbf{f}'_{e1} \cdot d\mathbf{s} \qquad (94d)$$

$$a_9 = \frac{1}{V}\int_{A_{12}} \nabla g'_{e0} \cdot d\mathbf{s} \qquad (94e)$$

$$\boldsymbol{a}_{10} = \frac{1}{V}\int_{A_{12}} \nabla \cdot \mathbf{g}'_{e1} \cdot d\mathbf{s} \qquad (94f)$$

$$a_{11} = \frac{1}{V}\int_{A_{12}} \nabla f'_{e0} \cdot d\mathbf{s} \qquad (94g)$$

$$\boldsymbol{a}_{12} = \frac{1}{V}\int_{A_{12}} \nabla \cdot \mathbf{f}'_{e1} \cdot d\mathbf{s} \qquad (94h)$$

The macroscopic equations for the potentials are

$$0 = \overline{\nabla}^2(\phi_1\overline{\boldsymbol{\Phi}}_1) - \boldsymbol{\Xi}_1 - \boldsymbol{\Psi}_1$$

$$= \overline{\nabla}^2(\phi_1\overline{\boldsymbol{\Phi}}_1) - (a_5 + a_9)(\overline{\boldsymbol{\Phi}}_1 - \overline{\boldsymbol{\Phi}}) - (\boldsymbol{a}_6 + \boldsymbol{a}_{10}) \cdot (\nabla\overline{\boldsymbol{\Phi}} - \sigma_{ei}\nabla\overline{\boldsymbol{\Phi}}_1) \qquad (99a)$$

$$0 = \overline{\nabla}^2(\phi\overline{\boldsymbol{\Phi}}_2) + \boldsymbol{\Xi}_2 + \boldsymbol{\Psi}_2$$

$$= \overline{\nabla}^2(\phi\overline{\boldsymbol{\Phi}}_2) + (a_7 + a_{11})(\overline{\boldsymbol{\Phi}}_1 - \overline{\boldsymbol{\Phi}}) - (\boldsymbol{a}_8 + \boldsymbol{a}_{12}) \cdot (\nabla\overline{\boldsymbol{\Phi}} - \sigma_{ei}\nabla\overline{\boldsymbol{\Phi}}_1) \qquad (99b)$$

Or we sum them up to be

$$\overline{\nabla}^2[(1-\phi)\overline{\boldsymbol{\Phi}}_1 + \phi\overline{\boldsymbol{\Phi}}_2] - \frac{1}{V}\int_{A_{12}} \nabla(\boldsymbol{\Phi}_1 - \boldsymbol{\Phi}_2) \cdot d\mathbf{s} - \frac{1}{V}\int_{A_{12}} (\boldsymbol{\Phi}_1 - \boldsymbol{\Phi}_2) \cdot d\mathbf{s} = 0 \qquad (100)$$

Note that at interface we get

$$\boldsymbol{\Phi}_1 - \boldsymbol{\Phi}_2 = \eta + \boldsymbol{\Phi}_d \qquad (101)$$

The intrinsic average of potential at the interface is

$$\bar{\boldsymbol{\Phi}}_1 - \bar{\boldsymbol{\Phi}}_2 = \bar{\eta} + \bar{\boldsymbol{\Phi}}_d \qquad (102)$$

Note that the overpotenital and potential of the electrochemical double layer only exit at the interface. Therefore, the macroscopic charge conservation for potential becomes:

$$\bar{\nabla}^2[(1-\phi)\bar{\boldsymbol{\Phi}}_1 + \phi\bar{\boldsymbol{\Phi}}_2] - \bar{\nabla}[\phi(\bar{\eta} + \bar{\boldsymbol{\Phi}}_d)] - [\phi(\bar{\eta} + \bar{\boldsymbol{\Phi}}_d)] = 0 \qquad (103)$$

The dispersions of Joule heat are

$$\bar{K}_1 = -\sigma_1\phi_1\overline{\nabla\boldsymbol{\Phi}_1' \cdot \nabla\boldsymbol{\Phi}_1'} = \sigma_1[a_1(\bar{\boldsymbol{\Phi}}_1 - \bar{\boldsymbol{\Phi}}) + \boldsymbol{a}_2 \cdot (\nabla\bar{\boldsymbol{\Phi}} - \sigma_{ei}\nabla\bar{\boldsymbol{\Phi}}_1)]^2 \qquad (104a)$$

$$\bar{K}_2 = -\sigma_2\phi\overline{\nabla\boldsymbol{\Phi}_2' \cdot \nabla\boldsymbol{\Phi}_2'} = \sigma_2[a_3(\bar{\boldsymbol{\Phi}}_1 - \bar{\boldsymbol{\Phi}}) + \boldsymbol{a}_4 \cdot (\nabla\bar{\boldsymbol{\Phi}} - \sigma_{ei}\nabla\bar{\boldsymbol{\Phi}}_1)]^2 \qquad (104b)$$

where $a_1$, $\boldsymbol{a}_2$, $a_3$ and $\boldsymbol{a}_4$ are closure coefficients.

8 Macroscopic governing equations with closure model

8.1 The macroscopic heat transfer equations for the integral electrolyte is

$$\rho_m C_p \frac{\partial(\phi\bar{T})}{\partial t} + \rho_m C_p \bar{\nabla} \cdot (\bar{T}\bar{\mathbf{u}}) = -k\bar{\nabla}^2(\phi\bar{T}) + \rho_m C_p \bar{\nabla} \cdot (\mathbf{A}_{\mathrm{D}} \cdot \bar{\nabla}\bar{T})$$

$$+k\bar{\nabla} \cdot [\boldsymbol{G}(\bar{\nabla}\bar{T} - \sigma_h\bar{\nabla}\bar{T}_1)] + h_{12}a_{12}(\bar{T}_1 - \bar{T}) + M_{12}a_{12} \cdot \bar{\mathrm{H}}_E + \frac{\sigma_2}{\phi}\bar{\nabla}(\phi\bar{\boldsymbol{\Phi}}_2) \cdot \bar{\nabla}(\phi\bar{\boldsymbol{\Phi}}_2) - \bar{K}_2 \qquad (105)$$

The macroscopic heat transfer equations for solid phase is

$$\rho_m C_p \frac{\partial(\phi_1\bar{T}_1)}{\partial t} = -k_1\bar{\nabla}^2(\phi_1\bar{T}_1) - k_1\bar{\nabla} \cdot [\boldsymbol{G}(\bar{\nabla}\bar{T} - \sigma_h\bar{\nabla}\bar{T}_1)] - h_{12}a_{12}(\bar{T}_1 - \bar{T})$$

$$-M_{12}a_{12} \cdot \bar{\mathrm{H}}_E + \frac{\sigma_1}{\phi_1}\bar{\nabla}(\phi_1\bar{\boldsymbol{\Phi}}_1) \cdot \bar{\nabla}(\phi_1\bar{\boldsymbol{\Phi}}_1) - \bar{K}_1 \qquad (106)$$

Therefore, for the integral electrolyte the macroscopic is shown in Table 1.

It should note that for the solvent, we only consider the mass conservation and momentum conservation since it is the matrix and inert to the electric force. Similarly, we can get the governing equations for the solvent

$$\overline{\nabla}\cdot(\phi\bar{\mathbf{u}}_s)=0 \qquad (107a)$$

$$\rho_m\left[\frac{\partial}{\partial t}(\phi\bar{\mathbf{u}}_s)+\overline{\nabla}\cdot(\phi(1+c)\bar{\mathbf{u}}_s\bar{\mathbf{u}}_s)\right]=-\overline{\nabla}(\phi\bar{p}_s)+\rho_{ms}\overline{\nabla}[(\nu+\chi)\overline{\nabla}\cdot(\phi\bar{\mathbf{u}}_s)]+\bar{\mathbf{b}}_s \qquad (107b)$$

Table 1 The macroscopic governing equations of the integral electrolyte with closure models

| Name | | Governing equation |
|---|---|---|
| Mass conservation | | $\overline{\nabla}\cdot(\phi\bar{\mathbf{u}})=0$ |
| Momentum conservation | | $\rho_m\left[\frac{\partial}{\partial t}(\phi\bar{\mathbf{u}})+\overline{\nabla}\cdot(\phi(1+c)\bar{\mathbf{u}}\bar{\mathbf{u}})\right]=-\overline{\nabla}(\phi\bar{p})+\rho_m\overline{\nabla}[(\nu+\chi)\overline{\nabla}\cdot(\phi\bar{\mathbf{u}})]+\bar{\mathbf{b}}_s$ |
| Energy equation | Fluid | $\rho_m C_p\frac{\partial(\phi\bar{T})}{\partial t}+\rho_m C_p\overline{\nabla}\cdot(\bar{T}\bar{\mathbf{u}})=-k\overline{\nabla}^2(\phi\bar{T})+\rho_m C_p\overline{\nabla}\cdot(\mathbf{A}_{\mathrm{D}}\cdot\overline{\nabla}\bar{T})$ $+k\overline{\nabla}\cdot[\boldsymbol{G}(\overline{\nabla}\bar{T}-\sigma\overline{\nabla}\bar{T}_1)]+h_{12}a_{12}(\bar{T}_1-\bar{T})+M_{12}a_{12}\cdot\overline{\mathrm{H}}_E+\frac{\sigma_2}{\phi}\overline{\nabla}(\phi\bar{\mathbf{\Phi}}_2)$ $\cdot\overline{\nabla}(\phi\bar{\mathbf{\Phi}}_2)-\bar{K}_2$ |
| | Solid | $\rho_m C_p\frac{\partial(\phi_1\bar{T}_1)}{\partial t}=-k_1\overline{\nabla}^2(\phi_1\bar{T}_1)-k_1\overline{\nabla}\cdot[\boldsymbol{G}(\overline{\nabla}\bar{T}-\sigma\overline{\nabla}\bar{T}_1)]-M_{12}a_{12}\cdot\overline{\mathrm{H}}_E$ $-h_{12}a_{12}(\bar{T}_1-\bar{T})+\frac{\sigma_1}{\phi_1}\overline{\nabla}(\phi_1\bar{\mathbf{\Phi}}_1)\cdot\overline{\nabla}(\phi_1\bar{\mathbf{\Phi}}_1)-\bar{K}_1$ |

8.2 The governing equations of species *i* with closure models

The macroscopic charge conservations are

$$\overline{\nabla}^2[(1-\phi)\overline{\mathbf{\Phi}}_1 + \phi\overline{\mathbf{\Phi}}_2] - \overline{\nabla}[\phi(\bar{\eta} + \overline{\mathbf{\Phi}}_d)] - [\phi(\bar{\eta} + \overline{\mathbf{\Phi}}_d)] = 0 \qquad (108)$$

$$\overline{\nabla}^2(\phi\overline{\mathbf{\Phi}}_d) = -\frac{1}{\varepsilon}\sum_i^n z_i F \mathbf{G}_{mi}\overline{\nabla}\bar{C}_i \qquad (109)$$

The mass conservation for ion is

$$\frac{\partial(\phi\bar{C}_i)}{\partial t} + \overline{\nabla}\cdot(\phi\bar{C}_i\bar{\mathbf{u}}_i) = \overline{\nabla}\cdot(\mathbf{M}_{\mathrm{D}i}\cdot\overline{\nabla}\bar{C}_i) + \dot{m}_{ci} \quad (110)$$

The momentum conservation for the ion is

$$-\overline{\nabla}(\phi\bar{p}) + \mu\overline{\nabla}\cdot(\phi\bar{\mathbf{S}}_i) + \bar{\mathbf{b}}_i - z_i F\bar{C}_i\overline{\nabla}(\phi\overline{\mathbf{\Phi}}_2) + z_i F\mathrm{F}_{\mathrm{D}i}\cdot\bar{C}_i\overline{\nabla}\overline{\mathbf{\Phi}}_2$$

$$-z_i F\boldsymbol{G}_{ei}(\overline{\nabla}\bar{C}_i - \sigma_{mi}\overline{\nabla}\bar{C}_{1i})\cdot(\nabla\overline{\mathbf{\Phi}} - \sigma_{mi}\nabla\overline{\mathbf{\Phi}}_1) = 0 \quad (111)$$

The concentration conservation in fluid is

$$\frac{\partial(\phi\bar{C}_i)}{\partial t} + \nabla\cdot(\phi\bar{C}_i\bar{\mathbf{u}}_i) = D_i\overline{\nabla}^2(\phi\bar{C}_i) + D_i\overline{\nabla}\cdot(\mathbf{G}_{mi}\overline{\nabla}\bar{C}_i)$$

$$+g_{12}a_{12}(C_i^0 - \bar{C}_i) + \gamma\overline{\nabla}\cdot[\bar{C}_i\overline{\nabla}(\phi\overline{\mathbf{\Phi}}_2)] + \gamma\overline{\nabla}\cdot[\boldsymbol{G}_{ei}(\overline{\nabla}\bar{C}_i - \sigma_{mi}\overline{\nabla}\bar{C}_{1i})\cdot(\nabla\overline{\mathbf{\Phi}} - \sigma_{mi}\nabla\overline{\mathbf{\Phi}}_1)]$$

$$+\gamma\boldsymbol{G}_{ei}(\overline{\nabla}\bar{C}_i - \sigma_{mi}\overline{\nabla}\bar{C}_{1i})\cdot(\nabla\overline{\mathbf{\Phi}} - \sigma_{mi}\nabla\overline{\mathbf{\Phi}}_1) - \overline{\nabla}\cdot(\mathbf{F}_{\mathrm{D}i}\cdot\bar{C}_i\overline{\nabla}\overline{\mathbf{\Phi}}_2) + \overline{\nabla}\cdot(\mathbf{M}_{\mathrm{D}i}\cdot\overline{\nabla}\bar{C}_i) \qquad (112)$$

The concentration conservation in solid is

$$\frac{\partial(\phi_1\bar{C}_{1i})}{\partial t} = \gamma_1\overline{\nabla}\cdot[\bar{C}_{1i}\overline{\nabla}(\phi\overline{\mathbf{\Phi}}_1)] - \gamma_1\overline{\nabla}\cdot[\boldsymbol{G}_{ei}(\overline{\nabla}\bar{C}_i - \sigma_{mi}\overline{\nabla}\bar{C}_{1i})\cdot(\nabla\overline{\mathbf{\Phi}} - \sigma_{mi}\nabla\overline{\mathbf{\Phi}}_1)]$$

$$+D_{1i}\overline{\nabla}^2(\phi_1\bar{C}_{1i}) - \gamma_1\boldsymbol{G}_{ei}(\overline{\nabla}\bar{C}_i - \sigma_{mi}\overline{\nabla}\bar{C}_{1i})\cdot(\nabla\overline{\mathbf{\Phi}} - \sigma_{mi}\nabla\overline{\mathbf{\Phi}}_1) - \overline{\nabla}\cdot(\mathbf{F}_{1\mathrm{D}i}\cdot\bar{C}_{1i}\overline{\nabla}\overline{\mathbf{\Phi}}_1)$$

$$-D_{1i}\overline{\nabla}\cdot(\mathbf{G}_{mi}\overline{\nabla}\bar{C}_i) - g_{12}a_{12}(C_i^0 - \bar{C}_i) + m_{ad} \qquad (113)$$

The current equation in the fluid is

$$\phi\bar{\mathbf{J}}_i = -z_i F[D_i\bar{\nabla}(\phi\bar{C}_i) + D_i\mathbf{G}_{mi}\bar{\nabla}\bar{C}_i + \gamma\bar{C}_i\bar{\nabla}(\phi\bar{\mathbf{\Phi}}_2) - \mathbf{F}_{\mathrm{D}i}\cdot\bar{C}_i\bar{\nabla}\bar{\mathbf{\Phi}}_2$$

$$+\gamma\boldsymbol{G}_{ei}(\bar{\nabla}\bar{C}_i - \sigma_{mi}\bar{\nabla}\bar{C}_{1i})\cdot(\nabla\bar{\mathbf{\Phi}} - \sigma_{mi}\nabla\bar{\mathbf{\Phi}}_1) - \bar{C}_i\bar{\mathbf{u}}_i + \mathbf{M}_{\mathrm{D}i}\cdot\bar{\nabla}\bar{C}_i] \qquad (114)$$

The current equation in the solid is

$$\phi_1\bar{\mathbf{J}}_{1i} = -z_i F[D_{1i}\bar{\nabla}(\phi_1\bar{C}_{1i}) - D_{1i}\mathbf{G}_{mi}\bar{\nabla}\bar{C}_i + \gamma_1\bar{C}_{1i}\bar{\nabla}(\phi\bar{\mathbf{\Phi}}_1) - \mathbf{F}_{1\mathrm{D}i}\cdot\bar{C}_{1i}\bar{\nabla}\bar{\mathbf{\Phi}}_1$$

$$-\gamma_1\boldsymbol{G}_{ei}(\bar{\nabla}\bar{C}_i - \sigma_{mi}\bar{\nabla}\bar{C}_{1i})\cdot(\nabla\bar{\mathbf{\Phi}} - \sigma_{mi}\nabla\bar{\mathbf{\Phi}}_1)] \qquad (115)$$

The macroscopic current equation of electron in solid is

$$\phi_1\bar{\mathbf{J}}_e = -\sigma_1\bar{\nabla}(\phi_1\bar{\mathbf{\Phi}}_1) - \sigma_1[a_5(\bar{\mathbf{\Phi}}_1 - \bar{\mathbf{\Phi}}) + \boldsymbol{a}_6\cdot(\nabla\bar{\mathbf{\Phi}} - \sigma_{ei}\nabla\bar{\mathbf{\Phi}}_1)] \qquad (116)$$

The macroscopic current equation at the interface by the electrochemical reaction is

$$\phi\bar{\mathbf{J}}_E = nF[k_f\phi\bar{C}_O - k_b\phi\bar{C}_R] \qquad (117)$$

The governing equations of ion with closure models are listed in Table 2.

Table 2 The macroscopic governing equations of species *i* with closure models

| Name | Equation |
|---|---|
| Charge conservation | $\bar{\nabla}^2(\phi\bar{\mathbf{\Phi}}_d) = -\frac{1}{\varepsilon}\sum_i^n z_i F\mathbf{G}_{mi}\bar{\nabla}\bar{C}_i$ <br> $\bar{\nabla}^2[(1-\phi)\bar{\mathbf{\Phi}}_1 + \phi\bar{\mathbf{\Phi}}_2] - \bar{\nabla}[\phi(\bar{\eta} + \bar{\mathbf{\Phi}}_d)] - [\phi(\bar{\eta} + \bar{\mathbf{\Phi}}_d)] = 0$ |
| Mass conservation | $\frac{\partial(\phi\bar{C}_i)}{\partial t} + \bar{\nabla}\cdot(\phi\bar{C}_i\bar{\mathbf{u}}_i) = \bar{\nabla}\cdot(\mathbf{M}_{\mathrm{D}i}\cdot\bar{\nabla}\bar{C}_i) + \dot{m}_{ci}$ |
| Momentum conservation | $-\bar{\nabla}(\phi\bar{p}) + \mu\bar{\nabla}\cdot(\phi\bar{\mathbf{S}}_i) + \bar{\mathbf{b}}_i - z_i F\bar{C}_i\bar{\nabla}(\phi\bar{\mathbf{\Phi}}_2) + z_i F\mathrm{F}_{\mathrm{D}i}\cdot\bar{C}_i\bar{\nabla}\bar{\mathbf{\Phi}}_2$ |

| | | |
|---|---|---|
| | | $-z_iF\boldsymbol{G}_{ei}(\bar{\nabla}\bar{C}_i - \sigma_{mi}\bar{\nabla}\bar{C}_{1i}) \cdot (\nabla\bar{\boldsymbol{\Phi}} - \sigma_{mi}\nabla\bar{\boldsymbol{\Phi}}_1) = 0$ |
| Concentration conservation | **Fluid** | $\frac{\partial(\phi\bar{C}_i)}{\partial t} + \nabla \cdot (\phi\bar{C}_i\bar{\mathbf{u}}_i) = D_i\bar{\nabla}^2(\phi\bar{C}_i) + D_i\bar{\nabla} \cdot (\mathbf{G}_{mi}\bar{\nabla}\bar{C}_i)$<br>$+g_{12}a_{12}(C_i^0 - \bar{C}_i) + \gamma\bar{\nabla} \cdot [\bar{C}_i\bar{\nabla}(\phi\bar{\boldsymbol{\Phi}}_2)] - \bar{\nabla} \cdot (\mathbf{F}_{\mathrm{D}i} \cdot \bar{C}_i\bar{\nabla}\bar{\boldsymbol{\Phi}}_2)$<br>$+\gamma\bar{\nabla} \cdot [\boldsymbol{G}_{ei}(\bar{\nabla}\bar{C}_i - \sigma_{mi}\bar{\nabla}\bar{C}_{1i}) \cdot (\nabla\bar{\boldsymbol{\Phi}} - \sigma_{mi}\nabla\bar{\boldsymbol{\Phi}}_1)]$<br>$+\gamma\boldsymbol{G}_{ei}(\bar{\nabla}\bar{C}_i - \sigma_{mi}\bar{\nabla}\bar{C}_{1i}) \cdot (\nabla\bar{\boldsymbol{\Phi}} - \sigma_{mi}\nabla\bar{\boldsymbol{\Phi}}_1) + \bar{\nabla} \cdot (\mathbf{M}_{\mathrm{D}i} \cdot \bar{\nabla}\bar{C}_i)$ |
| | **Solid** | $\frac{\partial(\phi_1\bar{C}_{1i})}{\partial t} = D_{1i}\bar{\nabla}^2(\phi_1\bar{C}_{1i}) + \gamma_1\bar{\nabla} \cdot [\bar{C}_{1i}\bar{\nabla}(\phi\bar{\boldsymbol{\Phi}}_1)]$<br>$-\gamma_1\bar{\nabla} \cdot [\boldsymbol{G}_{ei}(\bar{\nabla}\bar{C}_i - \sigma_{mi}\bar{\nabla}\bar{C}_{1i}) \cdot (\nabla\bar{\boldsymbol{\Phi}} - \sigma_{mi}\nabla\bar{\boldsymbol{\Phi}}_1)]$<br>$-\gamma_1\boldsymbol{G}_{ei}(\bar{\nabla}\bar{C}_i - \sigma_{mi}\bar{\nabla}\bar{C}_{1i}) \cdot (\nabla\bar{\boldsymbol{\Phi}} - \sigma_{mi}\nabla\bar{\boldsymbol{\Phi}}_1) - \bar{\nabla} \cdot (\mathbf{F}_{1\mathrm{D}i} \cdot \bar{C}_{1i}\bar{\nabla}\bar{\boldsymbol{\Phi}}_1)$<br>$-D_{1i}\bar{\nabla} \cdot (\mathbf{G}_{mi}\bar{\nabla}\bar{C}_i) - g_{12}a_{12}(C_i^0 - \bar{C}_i) + m_{ad}$ |
| Current equation | **Fluid** | $\phi\bar{\mathbf{J}}_i = -z_iF[D_i\bar{\nabla}(\phi\bar{C}_i) + D_i\mathbf{G}_{mi}\bar{\nabla}\bar{C}_i + \gamma\bar{C}_i\bar{\nabla}(\phi\bar{\boldsymbol{\Phi}}_2) - \mathbf{F}_{\mathrm{D}i} \cdot \bar{C}_i\bar{\nabla}\bar{\boldsymbol{\Phi}}_2$<br>$+\gamma\boldsymbol{G}_{ei}(\bar{\nabla}\bar{C}_i - \sigma_{mi}\bar{\nabla}\bar{C}_{1i}) \cdot (\nabla\bar{\boldsymbol{\Phi}} - \sigma_{mi}\nabla\bar{\boldsymbol{\Phi}}_1) - \bar{C}_i\bar{\mathbf{u}}_i + \mathbf{M}_{\mathrm{D}i} \cdot \bar{\nabla}\bar{C}_i]$ |
| | **Solid** | $\phi_1\bar{\mathbf{J}}_{1i} = -z_iF[D_{1i}\bar{\nabla}(\phi_1\bar{C}_{1i}) - D_{1i}\mathbf{G}_{mi}\bar{\nabla}\bar{C}_i + \gamma_1\bar{C}_{1i}\bar{\nabla}(\phi\bar{\boldsymbol{\Phi}}_1)$<br>$-\gamma_1\boldsymbol{G}_{ei}(\bar{\nabla}\bar{C}_i - \sigma_{mi}\bar{\nabla}\bar{C}_{1i}) \cdot (\nabla\bar{\boldsymbol{\Phi}} - \sigma_{mi}\nabla\bar{\boldsymbol{\Phi}}_1)] - \mathbf{F}_{1\mathrm{D}i} \cdot \bar{C}_{1i}\bar{\nabla}\bar{\boldsymbol{\Phi}}_1$ |
| | | $$\phi_1\bar{\mathbf{J}}_e = -\sigma_1\bar{\nabla}(\phi_1\bar{\mathbf{\Phi}}_1) - \sigma_1[a_5(\bar{\mathbf{\Phi}}_1 - \bar{\mathbf{\Phi}}) + \boldsymbol{a}_6 \cdot (\nabla\bar{\mathbf{\Phi}} - \sigma_{ei}\nabla\bar{\mathbf{\Phi}}_1)]$$ |
| | **Interface** | $$\phi\bar{\mathbf{J}}_E = nF[k_f\phi\bar{C}_O - k_b\phi\bar{C}_R]$$ |

8. Concluding remarks

In our previous work, the macroscopic governing equations are derived from the conservation laws in the macroscopic REV, the mixture of the solid and the fluid. At first, we define the porosity by the volume and surface and divide the porosity into various categories. Then the superficial averages is transformed into intrinsic averages to derive the interaction terms between the solid and the fluid, known as dispersion, tortuosity and interfacial transfer. The macroscopic governing equations are derived by performing the intrinsic average on the microscopic governing equations. After done that, the unknown terms related to the dispersion, tortuosity and interfacial transfer are emerged in the governing equations. In this paper we establish the closure models for these unknown terms to close the governing equations from mathematical algebra. The correlation between the stagnant and flowing coefficients of the fluid is also expressed. The coefficients appearing in the closure models have to be determined by the experiments or calculated by the numerical simulations [14-19]. Our new theory provides a new approach to the electrochemical flows-through porous media. It is suitable for the general case, in which the fluid comprising of solute and solvent flows through the porous media. But in the specific case, the equations could be further simplified based on reasonable assumption. In our next work, the Boltzmann distribution theory will be used to deal with the strong electrolyte or the mixture of gas with arbitrary components.

**Acknowledgements**

The author C. Xu would like to thank Dr. Sum Wai Chiang for helpful discussion. We like to thank the financial support from National Nature Science Foundation of China under Grants (No. 51102139) and from Shenzhen Technical Plan Projects (No. JC201105201100A). We also thank the financial support from Guangdong Province Innovation R&D Team Plan (2009010025). CT would like to thank City Key Laboratory of Thermal Management Engineering and Materials for financial support.

**Appendix**

Notation

REV representative elementary volume

mREV macroscopic representative elementary volume

sREV microscopic representative elementary volume

**w** physical quantities used for macroscopic representative elementary volume

$\bar{\mathbf{w}}$ average of a fluid quantity **w** over the fluid volume

*V* volume of macroscopic representative elementary volume

$\nabla$ Hamilton operator

$\nabla^2$ Laplacian operator

$\bar{\nabla}$ macroscopic gradient operator

$d_p$ particle size

$\rho_m$ mass density of fluid

$\rho_c$ total charge density

$z_i$ charge number of species *i*

$\rho_{mi}$ mass density of species *i*

$C_i$ concentration of species *i* in the fluid

$C_i^0$ concentration of species *i* in the bulk electrolyte

$C_{1i}$ concentration of species *i* in the solid

$M_i$ atomic weight of species *i*

*F* Faradic constant

$\rho_{ci}$ charge density of species *i*

**u** mean velocity of the fluid

$\bar{\mathbf{u}}$ intrinsic velocity of the fluid

$\mathbf{u}'$ dispersion of the velocity of the fluid

$\mathbf{u}_s$ mean velocity of the solvent group

*T* temperature

*S* surface area

***s*** unit surface

$D_i$ mass diffusion coefficient of species *i* in the fluid

$D_{sti}$ effective stagnant mass diffusion coefficient in the fluid for species *i* at local thermal equilibrium

$D_{1i}$ diffusion coefficient of species *i* in the solid phase

$D_{i1}$ mass dispersion diffusivity coefficient of species *i* in the longitudinal direction

$D_{i2}$ mass dispersion diffusivity coefficient of species *i* in the transverse direction

$D_{i3}$ mass dispersion diffusivity coefficient of species *i* in lateral direction

**S** strain rate tensor

$C_p$ thermal capacity of the fluid

*k* thermal conductivity of the fluid

$k_1$ thermal conductivity of thesolid

$k_{st}$ effective stagnant thermal conductivity of the fluid at local thermal equilibrium

$\bar{K}_1$ dispersion of the Joule heat generated inside the solid

$\bar{K}_2$ dispersion of the Joule heat generated inside the fluid

$p$ pressure

$T_1$ temperature of solid phase

$Re$ Reynolds number

$\mathrm{Re_p}$ local Reynolds number

$Pe$ Peclet number

$\mathrm{Pe_p}$ local Peclet number

$Pe_m$ Peclet number for the mass transfer

$Pe_{mp}$ local mass transfer Peclet number

$Pr$ Prandtl number

$Sc$ Schmidt number

$Sh$ Sherwood number

$Sh^*$ stagnant interfacial Sherwood number

$A_{12}$ fluid-solid interface area in macroscopic representative elementary volume

$O$ reactant in electrochemical reaction

$R$ product in electrochemical reaction

$C_O$ concentration of reactant $O$

$C_R$ concentration of product $R$

$C_0^O$ bulk concentration in the electrolyte of the reactant $O$

$C_0^R$ bulk concentration in the electrolyte of the product $R$

$n$ charge transfer number of the electrons in electrochemical reaction

$\mathrm{H}_E$ rate of heat per unit area generated by electrochemical reaction

$\mathrm{H}_s$ heat generated in the solid phase

$\mathrm{H}_r$ heat generated inside the fluid

$i$ current

$\mathbf{J}_i$ current flux of species $i$

$\mathbf{J}_0$ exchange current flux

$\mathbf{J}$ current flux

$\mathbf{J}_E$ neat reaction current per unit area

$\mathbf{J}_e$ current flux of electron in the solid phase

$\mathbf{J}_1$ current flux of ions in the solid phase

$\mathbf{J}_2$ current flux of ions in the fluid phase

$\overline{\mathbf{u}'\mathbf{u}'}$ momentum dispersion

$\bar{\mathbf{b}}$ interfacial force between the solid phase and the fluid phase of the electrolyte

$\overline{T'\mathbf{u}'}$ energy dispersion

$\Lambda_{12}$ thermal tortuosity

$\mathbf{q}_{12}$ interfacial heat transfer

$\overline{C_i'\mathbf{u}_i'}$ mass dispersion

$\mathbf{\Omega}_i$ mass transfer tortuosity of species $i$ at the interface

$\mathbf{\Xi}_1$ tortuosity of electric field in the solid

$\mathbf{\Xi}_2$ tortuosity of electric field in the fluid

$\mathbf{\Psi}_1$ tortuosity of potential in the solid

$\mathbf{\Psi}_2$ tortuosity of potential in the fluid

$m_{12}$ interfacial mass transfer from solid into fluid

$m_{ad}$ interfacial mass adsorption on the surface of the solid

$\bar{\mathbf{S}}_i$ stress rate tensor of species $i$

$\bar{\mathbf{b}}_i$ interfacial force of species $i$

$\mathbf{\Gamma}_i$ tortuosity of electric force of species $i$

$\overline{C'\nabla\mathbf{\Phi}'}$ dispersion of electric force

$\bar{l}$ mixing length of momentum dispersion

$c_1'$ closure coefficient in equation of velocity dispersion

$\mathbf{c}_2'$ closure coefficient in equation of velocity dispersion

$c$ closure coefficient in equation of momentum dispersion

$c_1$ closure coefficient in equation of momentum dispersion

$c_2$ closure coefficient in equation of momentum dispersion

$c_3$ closure coefficient in equation of momentum dispersion

$c_4$ closure coefficient in equation of interfacial heat transfer coefficient

$c_5$ closure coefficient in equation of interfacial mass transfer coefficient

$\mathrm{c_S}$ drag closure coefficient due to the Stoke's law

$\mathrm{c_B}$ viscous drag coefficient due to advection

$\mathrm{c_I}$ inviscid drag coefficient due to advection

$\mathrm{c_G}$ viscous lift coefficient due to advection

$\mathrm{c_L}$ inviscid lift coefficient due to advection

$\mathrm{c_M}$ inertia drag coefficient of Basset transient memory effect inertia drag coefficient

$\mathrm{c_V}$ virtual mass transient inertia drag coefficient

$a_i$ coefficient in equation of thermal diffusivity dispersion

$b_i$ coefficient in equation of thermal diffusivity dispersion

$a_{mi}$ coefficient in equation of mass dispersion diffusion coefficient

$b_{mi}$ coefficient in equation of mass dispersion diffusion coefficient

$a_1$ closure coefficient in equation of heat dispersion

$a_2$ closure coefficient in equation of heat dispersion

$a_3$ closure coefficient in equation of heat dispersion

$a_4$ closure coefficient in equation of heat dispersion

$a_5$ closure coefficient in equation of tortuosity of electric field

$\boldsymbol{a}_6$ closure coefficient in equation of tortuosity of electric field

$a_7$ closure coefficient in equation of tortuosity of electric field

$\boldsymbol{a}_8$ closure coefficient in equation of tortuosity of electric field

$a_9$ closure coefficient in equation of tortuosity of potential

$\boldsymbol{a}_{10}$ closure coefficient in equation of tortuosity of potential

$a_{11}$ closure coefficient in equation of tortuosity of potential

$\boldsymbol{a}_{12}$ closure coefficient in equation of tortuosity of potential

$h_{12}$ interfacial heat transfer coefficient

$M_{12}$ interfacial heat transfer coefficient

$h^*_{12}$ stagnant interfacial heat transfer coefficient

$g_{12}$ interfacial mass transfer coefficient

$g^*_{12}$ stagnant interfacial mass transfer coefficient

$a_{12}$ interfacial area density

$\mathbf{A}_D$ thermal dispersion diffusivity tensor

$\mathbf{M}_{Di}$ mass dispersion diffusion coefficient tensor of species *i* in the fluid

$F_{Di}$ electric force dispersion coefficient tensor of species *i* in the fluid

$\mathbf{F}_{Di}$ electric force dispersion coefficient tensor of species *i* in the fluid

$\mathbf{F}_{1Di}$ electric force dispersion coefficient tensor of species *i* in the solid

$\boldsymbol{G}$ thermal tortuosity parameter

$\boldsymbol{G}_{mi}$ mass tortuosity parameter of species *i*

$\boldsymbol{G}_{ei}$ tortuosity parameter of electric force for species *i*

Greek letters

ƙ Boltzmann constant

ꞷ constant in heat generated by the electrochemical reaction

$\varepsilon$ dielectric constant of the electrolyte

$\lambda_i$ diffusion conductivity

$\tau$ time scale of one electrochemical fluid

$\mathbf{\Phi}_1$ potential of solid phase and fluid

$\mathbf{\Phi}_2$ potential of electrochemical fluid

$\mathbf{\Phi}_d$ potential of the electrochemical double layer

$\mathbf{\Phi}_r$ potential of the representative elementary volume

$\mathbf{\Phi}_{ri}$ potential of ion species *i*

$\eta$ overpotential of electrochemical reaction

$\mu$ dynamic viscosity of the electrolyte

$\nu$ kinematic viscosity of the electrolyte

$\chi$ dispersion viscosity

$\alpha$ thermal diffusivity of the fluid

$\alpha_{st}$ stagnant thermal diffusivity of the fluid

$\alpha_1$ thermal dispersion diffusivity in the longitudinal direction

$\alpha_2$ thermal dispersion diffusivity in the transverse direction

$\alpha_3$ thermal dispersion diffusivity in the lateral direction

$\phi$ volume porosity of the fluid

$\phi_a$ area porosity of the fluid

$\phi_1$ volume porosity of the solid

$\sigma$ conductivity of the solid phase

$\sigma_h$ thermal conductivity ratio between solid and fluid

$\sigma_{mi}$ mass diffusion coefficient ratio between solid and fluid for ion species *i*

$\sigma_{ei}$ electric force coefficient ratio between solid and fluid for ion species *i*

$\gamma$ coefficients of the fluid in mass concentration conservation equation

$\gamma_1$ coefficient of the solid in mass concentration conservation equation

$\gamma_{st}$ effective stagnant coefficient

Subscripts

1 solid phase

2 fluid phase

*i* species

Superscripts

′ dispersion

Overhead bar

– intrinsic average of physical quantity